\documentclass[%
 aip,
 amsmath,amssymb,
 reprint,%
]{revtex4-1}

\usepackage{graphicx}
\usepackage{dcolumn}
\usepackage{bm}

\usepackage[utf8]{inputenc}
\usepackage[T1]{fontenc}
\usepackage{mathptmx}
\usepackage{etoolbox}
\usepackage{float}

\makeatletter
\def\@email#1#2{%
 \endgroup
 \patchcmd{\titleblock@produce}
  {\frontmatter@RRAPformat}
  {\frontmatter@RRAPformat{\produce@RRAP{*#1\href{mailto:#2}{#2}}}\frontmatter@RRAPformat}
  {}{}
}%
\makeatother
\begin{document}

\preprint{AIP/123-QED}

\title[Wipfli et al.]{Integration of a high finesse cryogenic build-up cavity with an ion trap}
\author{Oliver Wipfli}
 \author{Henry Fernandes Passagem}%
 \altaffiliation[]{hpassagem@phys.ethz.ch}
\affiliation{ 
Institute for Quantum Electronics, ETH Zürich, Otto-Stern Weg 1, 8093 Zürich, Switzerland
}%
\author{Christoph Fischer}%
\affiliation{ 
Institute for Quantum Electronics, ETH Zürich, Otto-Stern Weg 1, 8093 Zürich, Switzerland
}%
\author{Matt Grau}%
\affiliation{ 
Institute for Quantum Electronics, ETH Zürich, Otto-Stern Weg 1, 8093 Zürich, Switzerland
}%
\author{Jonathan P. Home}%
\altaffiliation[]{jhome@phys.ethz.ch}
\affiliation{ 
Institute for Quantum Electronics, ETH Zürich, Otto-Stern Weg 1, 8093 Zürich, Switzerland
}%

\date{\today}

\begin{abstract}

We report on the realization of a hemispherical optical cavity with a finesse of $F =$ 13000 sustaining inter-cavity powers of 10~kW, which we operate in a closed-cycle cryostat vacuum system close to 4 Kelvin. This was designed and built with an integrated radio-frequency Paul trap, in order to combine optical and radio-frequency trapping. The cavity provides a power build-up factor of 2250. We describe a number of aspects of the system design and operation, including low-vibration mounting and locking including thermal effects at high powers. Thermal self-locking in the high intracavity power regime was observed to enhance the passive stability below 1~kHz. Observations made over repeated cool-downs over a course of a year show a repeatable shift between the  ion trap center and the cavity mode. 

\end{abstract}

\maketitle


\section{\label{sec:introduction} Introduction}

The interest in operation of atomic and optical systems at cryogenic temperatures is growing, due to potential advantages with respect to vacuum conditions and the flexibility of materials which can be introduced \cite{spivey2021high, PhysRevApplied.16.034013}. Cryogenic vacuum systems also benefit from rapid cycle times for making changes to the in-vacuum elements, because they alleviate the need for  high-temperature baking of the vacuum system. This allows for rapid reconfiguration of the system. Alongside these benefits, however, cryostats come with additional complications compared to operation at room temperature, including sources of vibration from refrigeration, susceptibilty to high heat-loads, and the need to consider thermal contraction of materials. Since quantum-optical systems often rely on elements which are highly position sensitive at interferometric levels, the effects of these complications are severe. This is particularly the case for systems based on high-finesse cavities, which are extremely sensitive to fluctuations in the relative positions of the cavity mirrors. Despite this, cryogenic high finesse cavities have been previously realized in the context of optical frequency standards \cite{notcutt1995temperature, kessler2012sub, robinson2019crystalline, wiens2020simplified}, tests of relativity, and fundamental physics \cite{wiens2016resonator, PhysRevLett.88.010401}.

In this work, we present a high finesse Fabry-Perot cavity in a closed-cycle cryogenic vacuum system with a ring Paul trap surrounding the cavity mode. Mechanical vibrations in the cavity were reduced by suspending it using copper braids. Additionally, we investigated how the intracavity laser power changes the cavity resonances and affects the locking stability. We also characterize the cavity mode displaciment during the cooldown of the cryostat. 

The original motivation for the work described here was  quantum simulation, where it was envisioned that optical trapping could circumvent some of the challenges of operating Paul and Penning traps for 2-dimensional crystals of ions \cite{RevModPhys.75.281,cetina2012micromotion, bohnet2016quantum, mielenz2016arrays}. This goal followed from recent pioneering results on the optical trapping of magnesium and barium ions \cite{schneider2010optical, enderlein2012single, huber2014far, schaetz2017trapping, schmidt2018optical, karpa2021interactions}. The major challenge of the optical approach for ions spaced by distances on the order of a few tens of microns is that the Coulomb forces between ions are extremely strong compared to standard optical dipole forces produced using far detuned, internal state insensitive light. This places a premium on optical intensity. To meet this challenge, we planned to use a build-up optical resonator with high finesse. The deep optical potential (along with appropriately chosen atomic structure and laser detuning) formed potentially allows long term storage of ions, but is susceptible to ion loss through collisions with energetic atoms or molecules in the imperfect vacuum. For this reason it is highly desirable to operate such a system in as good a vacuum as possible. A cryogenic system was therefore chosen, since vacuum levels $\ll $10$^{-12}$~mbar have been reported \cite{micke2019closed, pagano2018cryogenic}.  

The presented work is relevant for other research fields that use optical cavities in a cryostat platform. Examples include investigation of interaction between photons and solid-state emitters\cite{vadia2021open} and frequency comb spectroscopy where an optical build-up resonator is used to increase the light-matter interaction path length \cite{grinin2020two}.

\section{\label{sec::Experimental Setup} Experimental setup}

The system aimed to position the trapping location of a radio-frequency Paul trap within the mode of a high finesse optical cavity with the capability to build up more than 10 kW of circulating power and a (1/$e{^2}$ intensity) beam radius of 110 $\mu$m at the ion trap position. The ion trap was used to trap magnesium ions, which have a relatively small differential AC Stark shift between the 3$^{2}$S$_{1/2}$ ($-$1.73 h Hz/(W/cm$^{2}$)) and 3$^{2}$P$_{3/2}$ ($-$1.57 h Hz/(W/cm$^{2}$)) levels for 1064~nm light.

A challenge of integrating the ion trap is that optical access to the centre of the trap is required for free space control beams as well as for imaging of the fluorescence of the atoms. This, along with the need to satisfy a maximum intensity due to damage at the mirror surfaces constrains the geometry of the build-up cavity, as well as the mounting structures for the trap, cavity mirrors and imaging objective.  

\subsection{\label{sec::Integrated Cavity-Ion trap in a Cryostat} Cavity and ion trap design}

\begin{figure}[h!]
\includegraphics[scale=0.8]{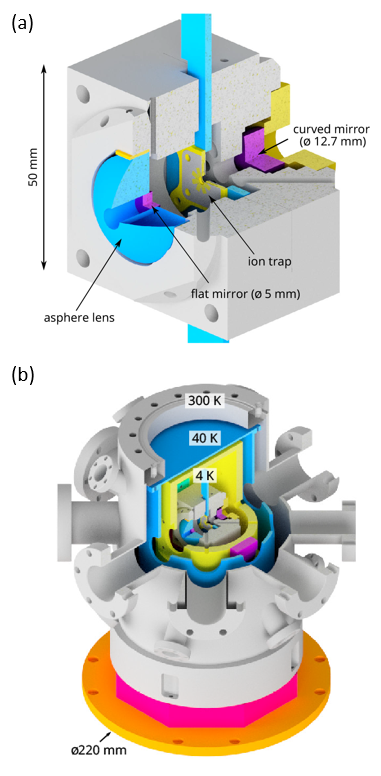}
\caption{\label{fig:cavity-cad} (a) A 29~mm long cavity is formed by a flat 5 mm diameter mirror and a concave 12.7 mm diameter mirror with a radius of curvature of 50 mm. The cavity beam goes through the center of a cylindrical ion trap with an RF electrode in the middle and two plates of segmented DC electrodes. This trap naturally forms 2d ion crystals that stand perpendicular to the cavity beam. We image such 2d ion crystals along the cavity axis with an asphere with a hole in the center. (b) Integrated cavity-ion trap mount, 4~K and 40~K shields in the stainless stell cryogenic vacuum chamber. 
}
\end{figure}

Our design for combining a hemispherical optical cavity with a ring ion trap in a closed-cycle cryogenic vacuum chamber is shown in Fig. \ref{fig:cavity-cad}. We chose the hemispherical geometry since it is robust to translation of the two mirrors relative to the others, helping to alleviate concerns that the cavity could become misaligned during the cooling down of the system.  The cavity length is $L$ = 29 mm, and the curved mirror has a radius of curvature of $R$ = 50 mm resulting in a stability parameter $g =$ $-$ 0.72. The mirror substrates were obtained from Perkins Precision Developments, and were specified to be polished to a surface roughness below 1.5 \AA \ and flatness of $\lambda/10$ at 633 nm. The substrates were then coated for high reflectivity at 1064 nm with a specified transmission of 200 ppm and total losses below 20 ppm. Although we initially considered using a piezo to control the position of one of the cavity mirrors, in practice the absolute frequency of our cavity is not critical, rather we prioritize stable power levels. For this reason we decided to fix both mirrors directly to the mount, and control the frequency of the light, rather than the length of the cavity, to keep the light resonant with the cavity.

Titanium grade 5 was used as the base material for the mount because it has a small coefficient of thermal expansion, is non-magnetic, and is stiff. The mount consists of two halves. The curved mirror is fixed to one half and  the aspheric lens holding the flat mirror is fixed to the other. The ion trap is mounted between the two halves. The mount has holes at 45$^{\circ}$ relative to the cavity axis for laser beam access as well as to allow neutral atoms from the oven to pass through the ion trap. 

The ion trap is formed by a stack of three electrode wafers located 16 mm away from the flat cavity mirror. Each wafer has a hole of 2~mm diameter in the center, and the total height of the stack is 1.35~mm. This allows access for laser beams which are directed at a 45 degree angle to the plane of the stack. The outer wafers of the stack are fabricated using electroplated gold on alumina, and are segmented to produce eight radial DC electrodes which equally divide up the circumference of the hole. These DC electrodes generate a static quadrupole potential, as well as allowing compensation of stray electric fields in the RF trap. The middle wafer of the stack is made of copper, and forms a metal ring to which radio-frequencies are applied for radio-frequency (RF) trapping. Typical RF voltages of 70~V peak-to-peak (pp) at 5.4~MHz are used to trap ions.

In order to allow imaging of Mg$^+$ (fluorescence wavelength of 280~nm) at the center of the cavity, the flat 5~mm diameter cavity mirror was mounted inside a central hole in a 25 mm diameter fused-silica aspheric lens with a focal length of $f$ = 17 mm. The lens working distance was 17 mm, and had a numerical aperture of NA = 0.46 including the 5~mm diameter hole in the middle.



\section{\label{sec:cavity-characterization} Cavity Characterization at Room Temperature}

\begin{figure}[h!]
\includegraphics[scale=0.58]{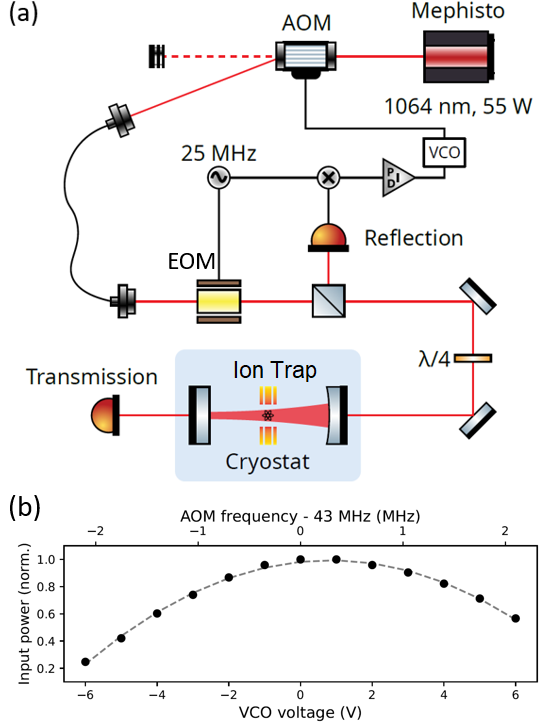}
\caption{\label{fig:optics-AOM} (a) Fast feedback setup using an AOM to lock the 1064 nm laser frequency to the optical cavity. An EOM added 25 MHz sidebands to the laser beam for the PDH error signal generation. (b) The cavity input power is AOM frequency dependent. The VCO voltage changes the AOM frequency with a tunability of 350 kHz/V.
}
\end{figure}

We characterize the cavity performance using 1064~nm laser light from a high power laser (Coherent model Mephisto MOPA 55 W). Frequency control of the light can be performed either with the  piezo of the laser, or using an acousto-optical modulator (AOM) with 40~MHz center frequency. To control the frequency of the AOM, we used a PID lock box (Vescent model D2-125) with a servo output voltage between -10 V and 10 V and a nominal bandwidth of 10~MHz, which is used to control a Minicircuits voltage-controlled oscillator (VCO) model ROS-43-119+. The laser beam passing through the AOM is deflected by a frequency-dependent angle and is subsequently coupled to an optical fiber delivering light to the cavity. Fig. \ref{fig:optics-AOM}(a) shows components of the cavity feedback system. To generate the Pound-Drever-Hall (PDH) error signal \cite{black2001introduction}, 25~MHz frequency sidebands were added to the 1064 nm laser beam using an electro-optic  modulator (EOM) model  EO-T25K3-NIR from Qubig. The carrier and sideband tones reflected off the cavity interfere on a photodiode. The signal from the photodiode is then mixed with a local oscillator, thus generating the error-signal for locking. An important element which is relevant to the  discussion later in this paper is the power change of the light incident to the cavity as the AOM is scanned. This dependence is shown  as function of VCO voltage in Fig. \ref{fig:optics-AOM}(b), normalized to the peak power. 

\begin{figure}[h!]
\includegraphics[scale=0.7]{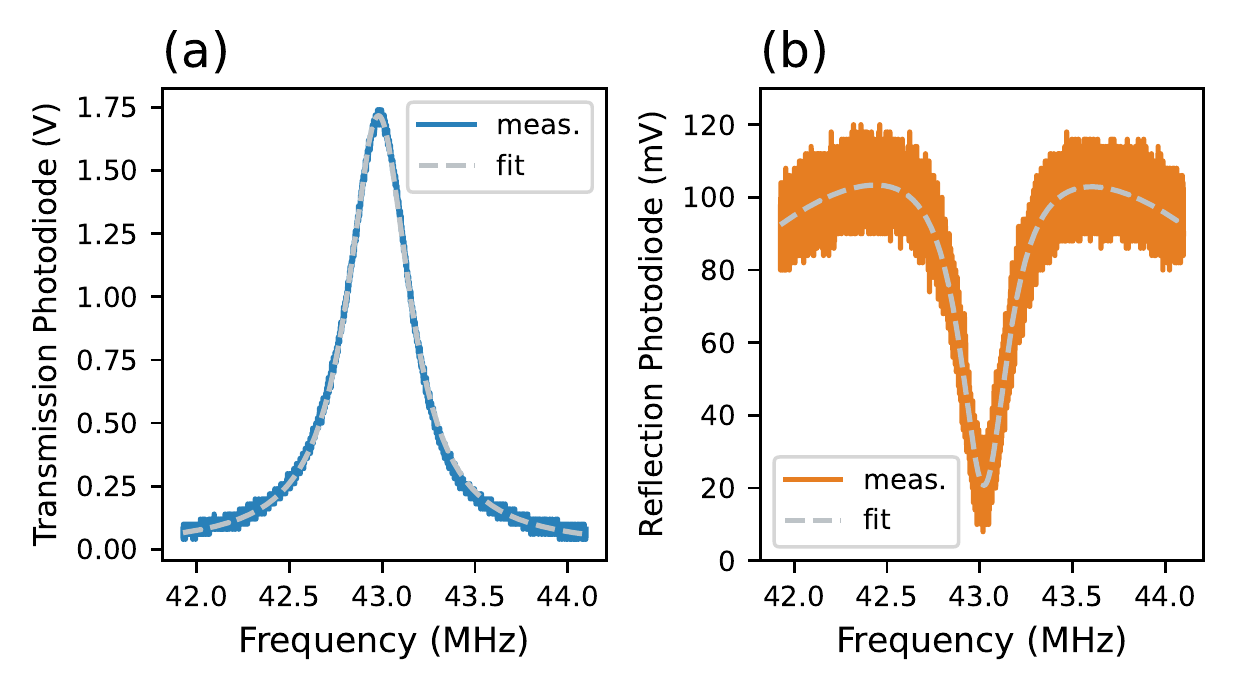}
\caption{\label{fig:cavity-characterization} Cavity characterization by sweeping the laser frequency with the AOM. (a) Transmission signal (blue) fitted by a Lorentzian profile (dashed line) extracting a cavity linewidth of $\kappa/2\pi = 403$~kHz. (b) Reflection signal (orange) fitted by an inverted Lorentzian curve superposed on a parabola, as the power after the AOM is  frequency-dependent and largest at 43~MHz.
}
\end{figure}

The first characterization of the optical cavity was performed at room temperature (300 K) with an input laser power of $P_{in} = 1.05$ mW. Fig. \ref{fig:cavity-characterization}(a) shows an example transmission profile when the laser frequency is swept over the cavity resonance using the AOM.
Its full width at half maximum (FWHM) was measured to be $\kappa/2\pi = 403$~kHz resulting in a length linewidth of $\delta_L =41$ pm at 1064~nm. The cavity length linewidth gives an indication of the tolerance of the cavity to cryostat vibration amplitudes. For passive stability, vibration amplitudes must be small compared to the cavity length linewidth, otherwise the cavity moves out of resonance with the laser.


Following the measurement proposed by Hood et al\cite{PhysRevA.64.033804}, we can estimate the intensity transmission coefficient $T$ and the loss coefficient $A$ of the cavity mirrors by measuring the transmitted power $P_{t}$ through the cavity on resonance, the power reflected $P_{r}$ on resonance and the input power $P_{in}$ which are related to the transmission, loss, and the intensity reflection co-efficient $R$ through
\begin{equation}
   \frac{P_{t}}{P_{r} - P_{in}} = \frac{T^2}{R A^2 - (T + A)^2} , 
\end{equation}
with
\begin{equation}
	R + T + A = 1
\end{equation}
and through a measurement of the cavity linewidth $\kappa$, which is related to the same parameters through
\begin{equation}
	\kappa = \frac{c(T+A)}{L} . 
\end{equation}
In our case $\kappa =2\pi \times$ 403~kHz, $L =$ 29~mm is the cavity length (obtained  from the mechanical design), $c$ is the speed of light. We assume that both cavity mirrors are identical since they were coated during the same coating run. The transmitted power of $P_{t} =$ 0.326~mW was measured while locking the cavity using the frequency feedback with the AOM. The reflected power $P_{r} =$ 0.189~mW on resonance was detemined by scanning the AOM frequency and recording the reflection photo-diode signal, see for example Fig. \ref{fig:cavity-characterization}(b).  This is fitted using a combination of a parabola superposed on an inverted Lorentzian which accounts for the  frequency-dependence of the AOM (Figure \ref{fig:optics-AOM}(b)). The fit is converted to on-resonance reflected power by comparison with the off-resonance reflected power  $P_{in} =$ 1.05~mW measured independently using a power meter. This set of measurements results in a per-mirror transmission coefficient $T =$ 135~ppm and loss coefficient $A =$ 110~ppm. The cavity finesse $F$ is then found to be  $F = \pi/(T + A) =$ 12820. The resonant power enhancement factor is given by $T/(T + A)^2 = 2250$. Table ~\ref{tab:cavity-characterization} summarizes the measured cavity parameters.

\begin{table}[H]
    \centering
    \begin{tabular}{l l}
Quantity &\qquad Value  \\
\hline
Cavity length $L$ &\qquad 29~mm  \\
Operating wavelength $\lambda$ &\qquad 1064~nm \\
Transmission $T$ &\qquad 135~ppm \\
Loss $A$ &\qquad 110~ppm \\
Finesse $F$ &\qquad 12820 \\
    \end{tabular}
    \caption{Summary of cavity parameters. Transmission $T $and loss $A$ are given as per-mirror intensity coefficients.}
    \label{tab:cavity-characterization}
\end{table}


\section{Cryogenic operation}

The geometry of the closed-cycle cryostat (Attocube model AttoDry 800) used in our experiment has its Gifford-McMahon head below the cold-plate of the cryostat, with the latter being in-plane with the surface of the optical table. This is convenient for assembling the cryogenic system. The cryostat model was chosen for low specified vibrations after comparing with models from competitors. Although the manufacturer supplies a vacuum chamber, due to various constraints of our planned experiments we built our own isolation vacuum chamber and 40~K shield. The vacuum sealing of the isolation vacuum chamber to the cryostat base-plate was performed via an O-ring supplied by the manufacturer. With typical heat-loads for running the ion trap in our experiment,  the temperature of the cryostat measured at the cavity mount was found to be 6~K. We did observe changes in temperature when higher optical powers were coupled to the cavity, these are detailed below. 

\subsection{Cavity mounting and vibration measurement}

\begin{figure}[h!]
\includegraphics[scale=0.95]{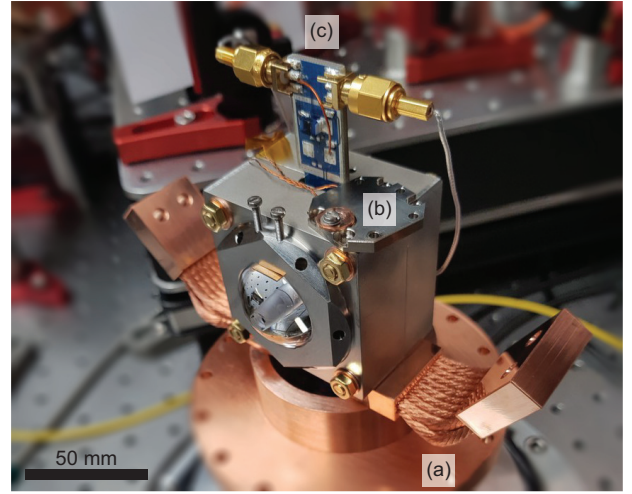} 
\caption{
\label{fig:Photo-cavity} The titanium cavity mount is suspended with copper braids (a) which damp vibrations from the closed-cycle croystat and conduct well thermally. (b) The cavity mount temperature is measured and stabilized by a resistive heater in the range of 6 to 11 K. (c) Coaxial cables connect to the trap printed-circuit-board (PCB) provide the 5.4 MHz drive tone.
}
\end{figure}

In our final mode of operation, the cavity block was suspended by copper braids to the 6-Kelvin parts of the cryostat. The ends of the wires forming the braids are clamped using a solderless, heat-free process in solid copper blocks featuring through-holes for mechanic mounting. The use of a high number of copper wires leads to a considerable cross-sectional area and high thermal conductivity. Thermal conductivity is critical because of the heat load presented by the radio-frequency drive to the RF trap as well as due to optical absorbtion at high optical powers. Vibrations from the cryostat propagate along the 4 K shield and have to pass through the braids to reach the cavity mount. The braids act as a spring-like system with a low resonance frequency (Hz level), and above resonance, they behave as a mechanical low-pass filter \cite{de2019vibration, ryou2017active, d2018active}. Fig. \ref{fig:Photo-cavity} shows how the copper braids are fixed to the integrated cavity and RF trap titanium mount.

\begin{figure}[h!]
\includegraphics[scale=0.9]{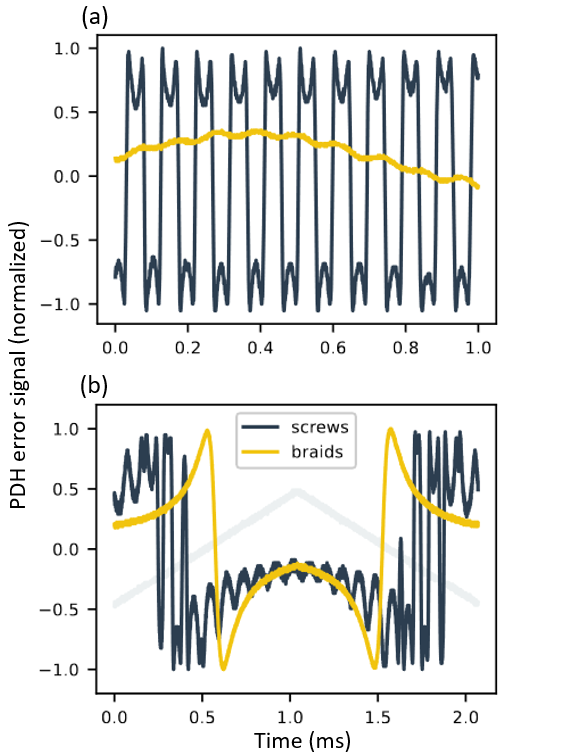}
\caption{\label{fig:time-traces-paper} Normalized PDH error signal measurements for (a) free-running laser frequency (constant frequency) and (b) scanning the laser frequency up and down through the resonance (indicated by the gray ramp). The minimum and maximum of the error signal are separated in frequency by a cavity linewidth of $\kappa/2\pi = 400$~kHz.  The black traces show the vibrations, due to strong mechanical coupling when the cavity mount was screwed to the cryostat cold head.  The yellow traces show results with the improved  mechanical decoupling with the cavity block suspended by copper braids.    
}
\end{figure}

In our initial designs, we had hoped to avoid the use of such braids, because of concerns that they would lead to slow drifts in the cavity block position over time. The initial mounting system therefore consisted of screwing the cavity block directly to the cold plate of the cryostat. The design tried to make use of symmetric mounting, in the hope that this would lead to common-mode vibration over the cavity block and thus not affect the mirror separation. This was unsuccessful, even with various adaptions such as reducing the number of screws and adding significant mass load to reduce accelerations due to the vibration cycle of the cold head. Measurements of amplitude and frequency of vibrations were performed by observing the PDH error signal. Two types of measurements were performed. In one, the laser was at constant frequency (free-running), while in the other, the laser frequency was scanned linearly over the cavity resonance (scanning). Results are shown in Fig. \ref{fig:time-traces-paper}. In the absence of vibrations, the free-running PDH error signal would be constant, and the scanning signal would show a single cavity resonance. With vibrations present, the free-running error signal oscillates at the vibration frequency, and the scanning error signal shows multiple resonances, as the cavity moves in and out of resonance with the laser. Fig. \ref{fig:time-traces-paper} compares the PDH error signals when the cavity was screwed directly onto the cold head (black traces) versus suspending the cavity mount with copper braids (yellow traces). The error signals were normalized such that the minimal and maximal values would be -1 and +1, respectively. At these values, the cavity-laser detuning is approximately half a linewidth, $\kappa$/2. The data shown in this figure was taken shortly after a cryostat kick (which occurs once every second in the cycle of the cryo-cooler), at the moment when vibrations are at their strongest.

When the cavity mount was fixed with screws onto the cryostat cold head, the vibration amplitude was larger than the cavity linewidth. The oscillations reach an error signal value of $\pm$1, which corresponds to peak-to-peak excursions of a full cavity linewidth of $\kappa/2\pi = 400$ kHz. This means that the cavity length makes excursions of $\delta L = L\kappa/\omega_{c} = 41$~pm-pp. In the 1~ms long recording window, we count roughly 10 oscillations, with a vibration frequency of approximately 10~kHz. To compensate for these vibrations, the lock would have to change the laser frequency by more than 400~kHz-pp at a frequency of 10 kHz. When the cavity mount was suspended by the copper braids, the 10 kHz vibration frequency mode was damped significantly. We found that the vibration amplitude was 8 kHz-pp, which was about 2 percent of a cavity linewidth. In terms of cavity length displacement, the peak-to-peak vibration amplitude in length was 0.02$L\kappa/\omega_{c} =$ 0.8~pm. Thus the fractional length deviations are 0.8~pm / 29~mm = 2.8 $\times 10^{-10}$. 
This was more than a factor of 50 reduction compared to the direct mounting.

To analyze vibrations in frequency space, we measured the cavity-laser detuning via the PDH error signal. The error signal was used as a proxy for the cavity length change. The laser frequency was tuned at the cavity resonance and kept constant to produce a vibration power spectral density. The PDH error signal was recorded for 2 seconds, capturing approximately 2 cryostat cycles, with a sampling frequency of 2.5~MHz. The error signal was normalized such that the minimum and maximum values were -1 and +1, respectively, corresponding to a detuning of half a cavity linewidth. 

\begin{figure}[H]
\includegraphics[scale=0.26]{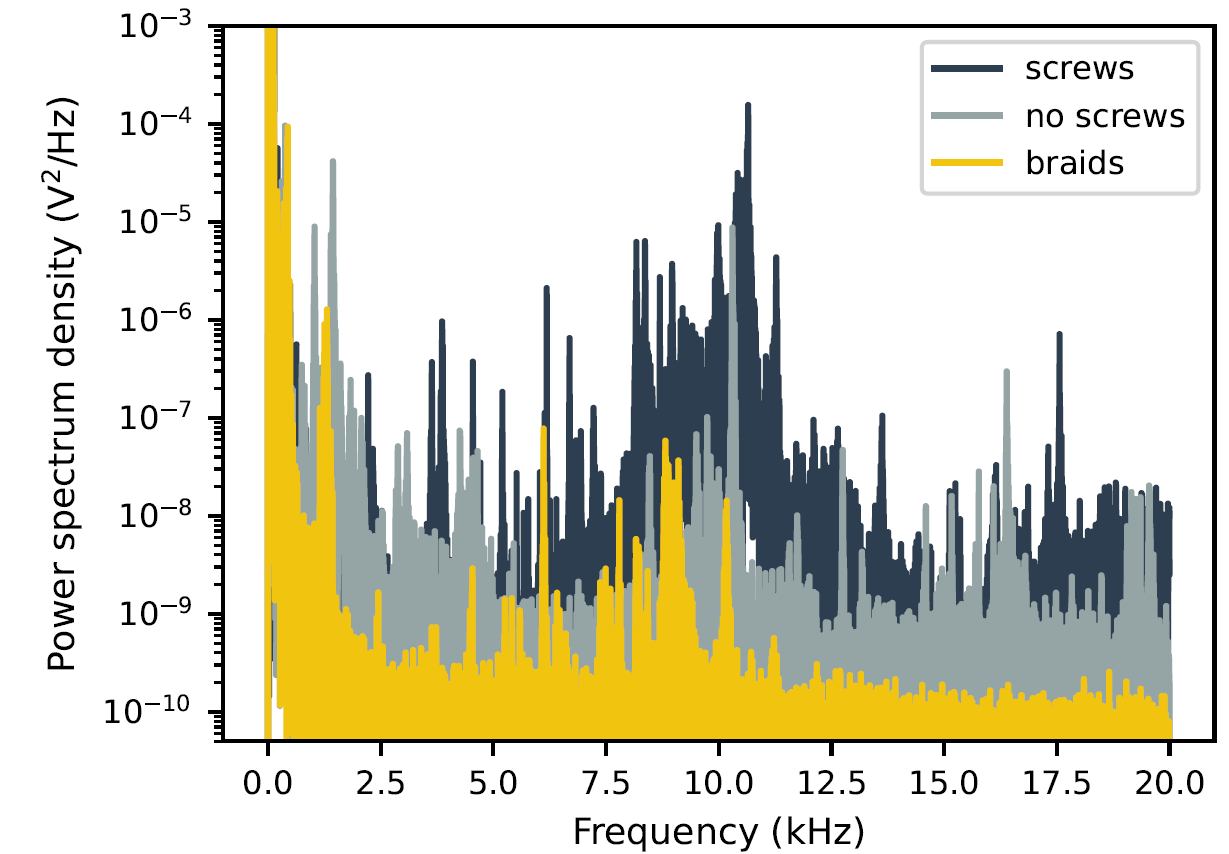}
\caption{\label{fig:power-spectra} Vibration power spectra density for three different mounting types: (screws) the cavity mount was fixed by screws on the cryostat cold head, (no screws) the cavity mount was only resting in place on the cryostat cold head, (braids) the cavity mount was suspended by the copper braids. The braids decoupled the cavity mechanically from the cold head and damped vibrations, mainly around 10 kHz. At frequencies below 1 kHz, the different mounting types performed similarly. The spectra have a frequency interval of $df =$ 0.5~Hz and acquisition bandwidth of $B_{w} =$ 2.5~MHz/2.56 = 977~kHz.
}
\end{figure}

Fig. \ref{fig:power-spectra} shows power spectral densities for the cavity mount assembled in the cryostat in three different ways: the cavity fixed on the cold head with screws (screws), the cavity resting on the cold head only without screws (no screws), and the cavity suspended using the copper braids and not touching the vibrating 4 Kelvin parts directly (braids). The power spectral density traces show that vibration amplitudes of frequency below 1 kHz were similar between the three mounting types. However, at higher frequencies, and especially around 10 kHz, the braids significantly reduced the vibration amplitude. The passive frequency stability improvement around 10 kHz by suspending the cavity with copper braids made the cavity locking possible when operating the cryostat. By contrast, we were unable lock the cavity without the copper braids. The initial concerns about drift of the braid-mounted cavity block position over time proved unfounded - the position was stable such that for both cavity coupling and for free-space beams with focusing lenses placed on the optical chamber outside the cryostat-setup no additional adaption had to be performed in day-to-day use.

\subsection{Laser power effects in the cavity}

\subsubsection{Scan Asymmetries and Lock Stability Conditions}

Circulating laser power in the cavity, combined with scattering and absorption losses, leads to power dissipation and subsequent temperature increases of the cavity mirrors and the materials surrounding them. Thermal expansion and contraction processes triggered by the change in the temperature lead to cavity length changes \cite{white1973thermal, PhysRevE.73.026217, konthasinghe2018dynamics}. These mechanical changes mean that in turn the cavity resonance frequency changes, which in turn affects the circulating power. Such thermal feedback processes depend on material properties like the thermal expansion coefficient, heat capacity and conductivity, but also on the cavity mounting geometry and the cavity spatial mode.

To describe the dynamics of thermal cavity length changes, 
we considered a simple model to express the cavity length as function of time:
\begin{equation}
    L(t) = L_{0}+ L_{s}(t) + L_{f}(t) ,
    \label{eq:Lt}
\end{equation}
\noindent where $L_{0}$ is a non-thermal cavity length (corresponding to the length in the absence of any mirror heating) , and the terms $L_{s}(t)$ and $ L_{f}(t)$ can be associated, respectively, to a ``slow'' photothermal expansion and a ``fast'' photothermal refraction processes \cite{konthasinghe2018dynamics}.
The equations of motion for the thermal cavity length changes are then
\begin{eqnarray}
\dot L_{s}(t) &= -\frac{1}{\tau_{s}}(L_{s}(t) + B_{s} P_{circ}(t)) \\
\dot L_{f}(t) &= -\frac{1}{\tau_{f}}(L_{f}(t) + B_{f} P_{circ}(t))
\end{eqnarray}
where $\tau_{s}$ and $\tau_{f}$ are the characteristic time scales for length changes, $B_{s}$ and $B_{f}$ are the length change per unit circulating power for the photothermal expansion and refraction processes, and $P_{circ}$ is the intracavity circulating power. In the approximation that length changes are much slower than the ring-down time of the cavity, we can assume that the circulating power depends on the cavity length, and thereby on the thermal length change \cite{lawrence1999dynamic} as
\begin{equation}
	P_{circ}(t) = \frac{\alpha P_{0}}{1+(\frac{4F}{\lambda}(L(t)-L_{res})^2)} ,
	\label{eq:Pcirc}
\end{equation}
where $P_{0}$ is the input power, $\alpha$ is the power build-up factor accounting for mode matching and impedance matching, $F$ is the cavity finesse, $\lambda$ is the laser wavelength, and $L_{res} = m \lambda /2$, with $m$ an integer, is the resonance length, i.e., the cavity length at which the cavity resonance frequency matches the laser frequency.

\begin{figure}[H]
\includegraphics[scale=0.76]{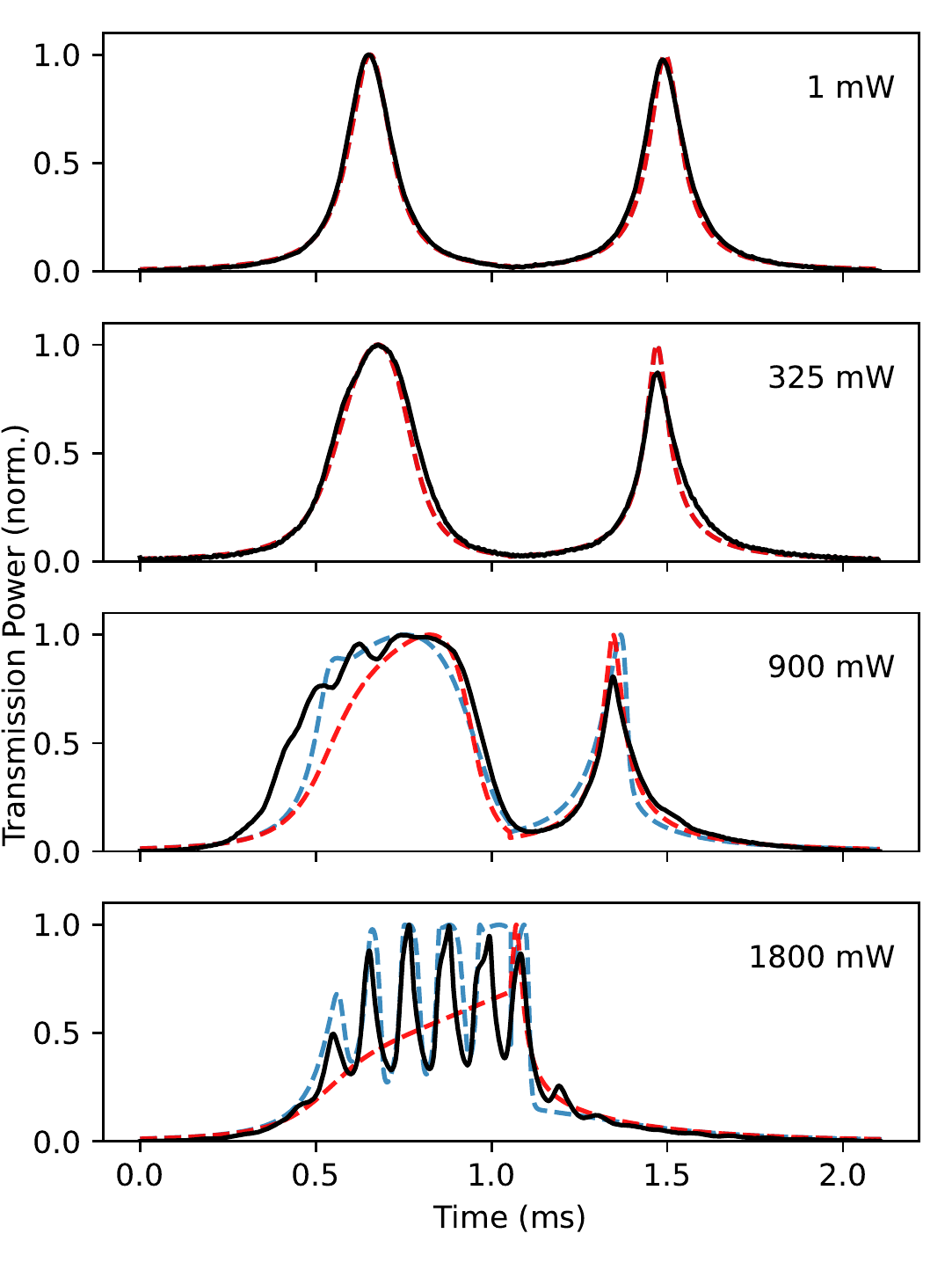}
\caption{\label{fig:thermal-scans-experiment&theory} 
Measurements of the transmitted power as the laser frequency is scanned linearly over a cavity resonance (black solid traces) for four different input powers. The peak transmitted powers are shown in the top right corner of each plot. The red dashed traces represent the thermal model for with only the photothermal expansion process included for $F = 12000$, $\alpha = 2000$, $B_{s} = 0.11$ pm/W and $\tau_{s} = 500$~$\mu$s. The blue dashed traces are the thermal model with both photothermal expansion and refraction processes included for $B_{f} = -14$~fm/W and $\tau_{f} = 5$~$\mu$s.
}
\end{figure}

Fig. \ref{fig:thermal-scans-experiment&theory} shows the measured transmitted power (solid black traces) as a function of time as the laser frequency was swept over the cavity resonance. The traces were recorded with a photodiode for different peak transmitted powers - these values are indicated at the top right corner of each plot. From 0-1~ms the laser frequency is decreasing linearly, while from 1-2~ms the laser frequency is increasing. At low power (1 mW peak transmission), the  lineshape is symmetric and independent of the scan direction.  At higher powers, we observe differences in the lineshape depending on the scan direction.  At powers of 900~mW and above, we additionally observe oscillatory behaviour in the broadened peak.

The lineshape asymmetry can be interpreted as follows. The scan starts with the laser wavelength below the cavity resonance, and as the laser wavelength increases and moves closer to the cavity resonance length $L_{res}$, the thermal length change $L_{s}(t)$ also increases. Since the scan is in the same direction as the resulting thermal change, the laser remains on resonance for longer, broadening the left peak. On the other hand, when the laser wavelength is scanned towards higher frequency, the cavity length still expands close to the resonance but the resonance and laser frequencies go in opposite directions, narrowing the right peak. 

To examine this interpretation, we modelled the system by numerically integrating equations \ref{eq:Lt} to \ref{eq:Pcirc}. Initially we included only the slower thermal cavity length change $L_s(t)$, and varied parameters manually to obtain a good match. The red dashed curve shows the results for $F = 12000$, $\alpha = 2000$, $B_{s} = 0.11$~pm / W, and $\tau_{s} = 500$ $\mu$s. This model reproduces the observed scan asymmetries, but does not produce oscillatory behaviour. 

The origin of the oscillatory effects can be qualitatively reproduced using a model which includes both a photothermal expansion and a refraction process acting on a different timescale with the opposite sign. For instance, including a second fast response $B_{f} = -14$~fm/W, $\tau_{f} = 5$~$\mu$s produces the curves shown in the dashed-blue traces in figure \ref{fig:thermal-scans-experiment&theory}. However we are not confident that we can ascribe the behaviour to an additional timescale in the thermal length response. The period of the observed oscillations was from 90 $\mu$s to 120 $\mu$s (roughly 9 kHz), coinciding with mechanical resonances of the cavity mount. The frequency of the observed oscillations was independent of the laser input power. An alternative explanation is therefore that high input laser powers excite mechanical self-resonances. This has been previously observed in a Fabry-Perot microcavity \cite{brachmann2016photothermal}. 

The thermal effects also affect the ability to lock the cavity using the AOM, since due to the response of the AOM the power input to the cavity changes with AOM frequency. At low powers, thermal effects were negligible and the lock was robust for any AOM frequency. However we noticed that at high cavity input powers (more than 100~mW), the lock only worked in a range of AOM frequencies where the cavity input power increases with increasing AOM frequency. In this scenario, the thermal shift due to the increase in input power results in a thermal cavity length change that reduces the detuning between the laser and cavity and thus helps the lock. In the other regime, a run-away process is activated. At the highest circulating powers, the thermal resonance frequency changes due to the non-constant AOM power profile dominate the detuning between laser and cavity, meaning that stable lock operation was only possible for negative VCO voltages. We found that the average VCO voltage had to be -2 V or less to  avoid run-away behavior, which meant that over its full range of -2 V$\pm$ 2 V the sign of the response helped the lock. Similar logic, but associated now with the response of the cavity to the input light, meant that the laser could not be locked to the peak of the cavity transmission. Thermal length changes were only stable on one side of the resonance, while on the other, instability prevailed. The run-away effects occurred on millisecond timescales, and they were avoidable with adequate lock settings. On minute timescales, we observed further cavity length changes and thermal self-locking.

\subsubsection{Cavity length change on minute timescales and thermal self-locking}

In addition to relatively fast changes in the cavity length which affect the locking, operating at high intracavity powers results in a rise in temperature of the materials surrounding the cavity over longer timescales. Over many minutes, we find that the cavity resonance frequency shifts by more than the AOM lock range of 1.5~MHz. To ensure that the laser did not come unlocked, a second lock was implemented where the laser frequency was tuned via the laser crystal temperature. This lock kept the average VCO voltage at a given setpoint and had a bandwidth of 1~Hz. 

\begin{figure}[H]
\includegraphics[scale=0.63]{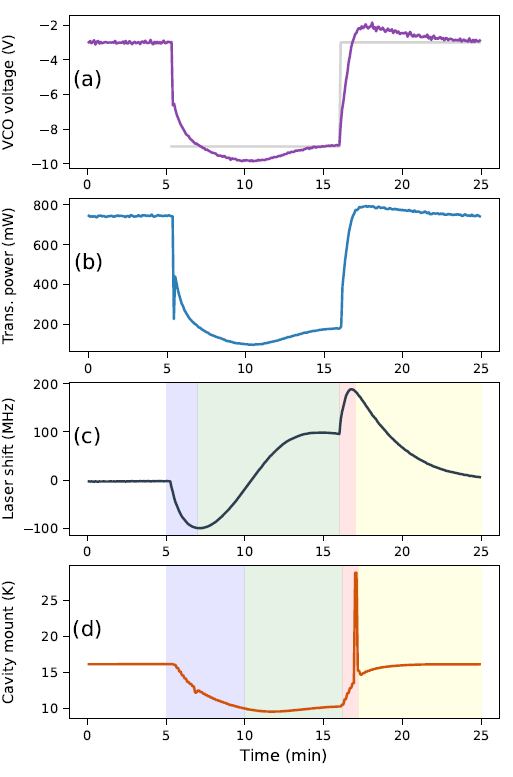}
\caption{\label{fig:1064-temperature} 
Thermal length changes over periods of 25 minutes. (a) VCO voltage (purple trace) and the chosen setpoint (gray trace). (b) Transmitted laser power through the cavity. This is closely related to the VCO voltage value due the AOM power profile. (c) Laser frequency is tuned to keep the VCO voltage at its setpoint. (d) Cavity mount temperature as a function of time.
}
\end{figure}

By sudden changes in the AOM frequency (controlled using the VCO) while the laser was locked, we were able to  rapidly change the input power to the cavity and thus observe the resulting changes in the mechanical lengths of the cavity by observing the voltage controlling the laser. In Fig. \ref{fig:1064-temperature} (a) and (b) we show the VCO voltage and the transmitted cavity power, respectively, recorded over a 25 minute period, during which the VCO setpoint was suddenly changed twice. Fig. \ref{fig:1064-temperature} (c) shows the laser frequency shift over time required to keep the VCO voltage close to the chosen setpoint, and thus is an indicator of the thermal extension of the cavity (the response of the second-stage lock is much faster than the timescales shown in the figure). Fig. \ref{fig:1064-temperature} (d) presents the temperature of the cavity mount measured over the corresponding period using a temperature sensor ((b) in Figure \ref{fig:Photo-cavity}). The first switch of the VCO setpoint occurs at minute 5, from $-3$~V (high power) to $-9$~V (low power). As a result, the cavity mount temperature decreased between minutes 5 and 10 (blue region in Fig. \ref{fig:1064-temperature} (d)) and settled to 11~K between minutes 10 to 15 (green region in Fig. \ref{fig:1064-temperature} (d)). This decrease in temperature was expected since the cavity input power was lowered. The laser frequency shift shows multiple thermal effects with different timescales and signs. Between minutes 5 and 7 it shifts by 100~MHz (blue region in Fig. \ref{fig:1064-temperature} (c)), but a slower but stronger thermal effect of opposite sign later starts to dominate, corresponding to a shift between $-$100~MHz and 100~MHz occuring between minutes 7 and 12. It seems to reach equilibrium between minutes 12 and 15 (green region in Fig. \ref{fig:1064-temperature} (c)). At minute 16, we again changed the VCO setpoint, but from $-9$~V (low power) to $-3$~V (high power). The temperature of the cavity mount then started increasing smoothly, but we suddenly observed a temperature spike at minute 17 (red region in Fig. \ref{fig:1064-temperature} (d)). Its source is unclear, but we speculate that thermal expansion might have changed the mechanical contact between parts of the setup. After the temperature spike, the cavity mount temperature settled at 16~K (yellow region in Fig. \ref{fig:1064-temperature} (d)). Between minutes 16 and 17, before the temperature spike, the laser frequency shift moved from 100~MHz to 180~MHz (red region in Fig. \ref{fig:1064-temperature} (c)). After minute 17 (temperature spike), the laser frequency decreased -180~MHz for the next eight minutes returning to its initial value (yellow region in Fig. \ref{fig:1064-temperature} (c)).  A cavity resonance frequency change of 100~MHz corresponds to a relative frequency change of 100~MHz/ (c/ 1064~nm) $\approx$ 3.3 $\times$ 10$^{-7}$; on the 29~mm cavity length of our cavity, this corresponds to $\approx$ 10.7~\rm nm.

\begin{figure}
\includegraphics[scale=0.57]{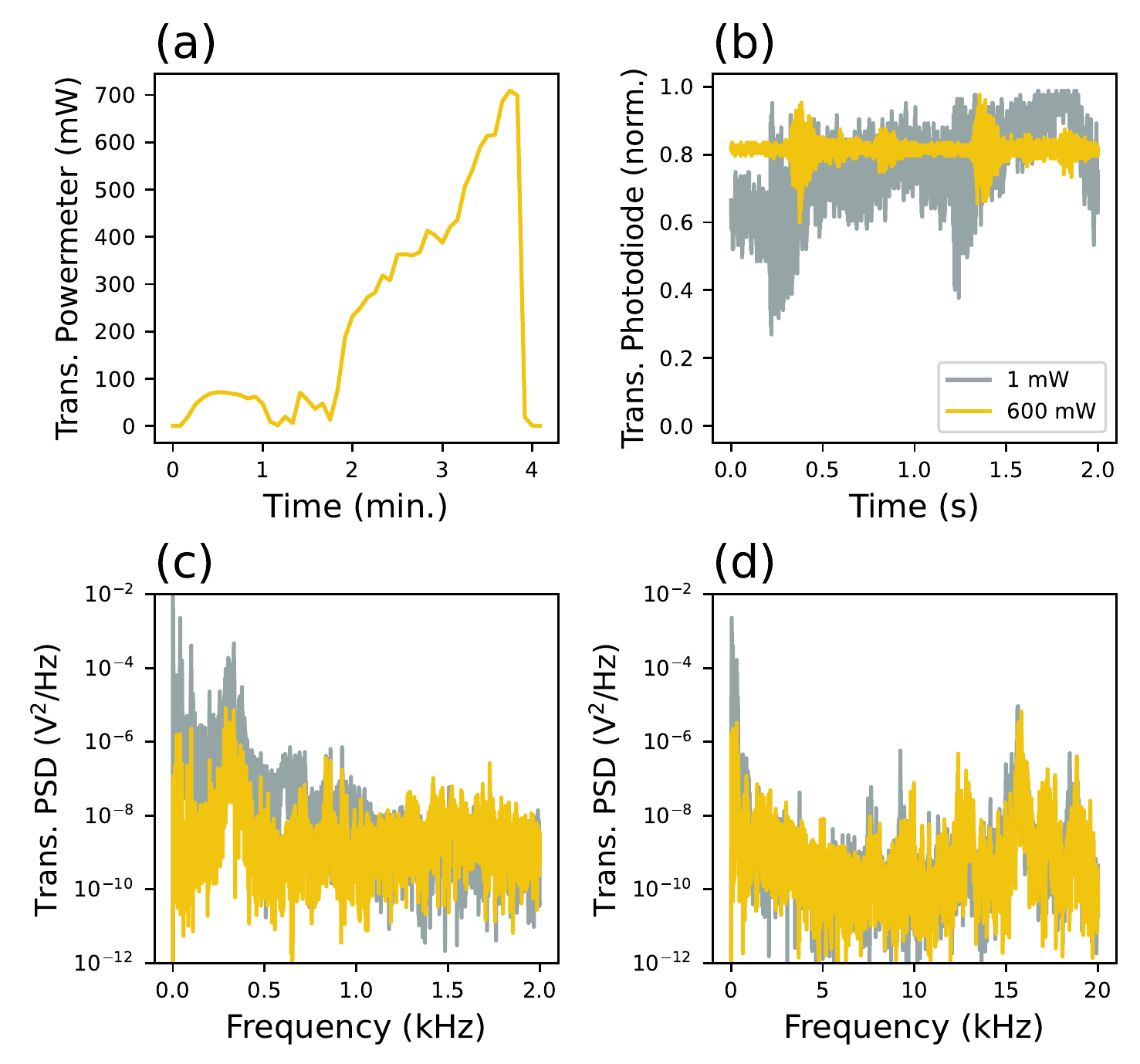}
\caption{\label{fig:self-locking} 
Thermal self-locking of the cavity resonance to a fixed laser frequency. Only the temperature of the cavity mount was actively stabilized with a heater. (a) The transmitted power was high over several minutes. (b) Transmitted power recorded over 2 seconds for the cases of low transmitted power (1~mW, gray) and high transmitted power (600~mW, yellow). (c) and (d) are the power spectral density (PSD) of the cavity transmission with a frequency interval of 0.5~Hz and acquisition bandwidth of 976~kHz. Noise was suppressed below 1 kHz, while at higher frequencies, no significant effect was observed.
}
\end{figure}

By stabilizing the cavity mount temperature using a resistive heater and with all active feedback on the laser turned off, we observed spontaneous thermal self-locking. Fig. \ref{fig:self-locking} (a) shows that the transmitted power was high over several minutes without any feedback on the laser frequency. To estimate the noise-suppression capabilities of this thermal self-locking process, we measured the transmitted power with a photodiode over a time of 2 seconds (Fig. \ref{fig:self-locking} (b)) and calculated the power spectral density (PSD) of the cavity transmission signals (Figure \ref{fig:self-locking} (c) and (d)). The measurements were performed at low circulating power with negligible thermal effects (gray traces) and high circulating power (yellow traces). We found that at frequencies below 1~kHz, the thermal self-locking removed noise, while at higher frequencies, no suppression was observed. This is in agreement with the estimated thermal time constant $\tau_{s} =$ 500~$\mu$s in the lineshape asymmetry measurements. The thermally self-locked signal achieves a length stability of 0.6~pm~rms which corresponds to a frequency fluctuation of 6~kHz~rms.

\subsubsection{Finesse Preservation}

\begin{figure}[H]
\includegraphics[scale=0.58]{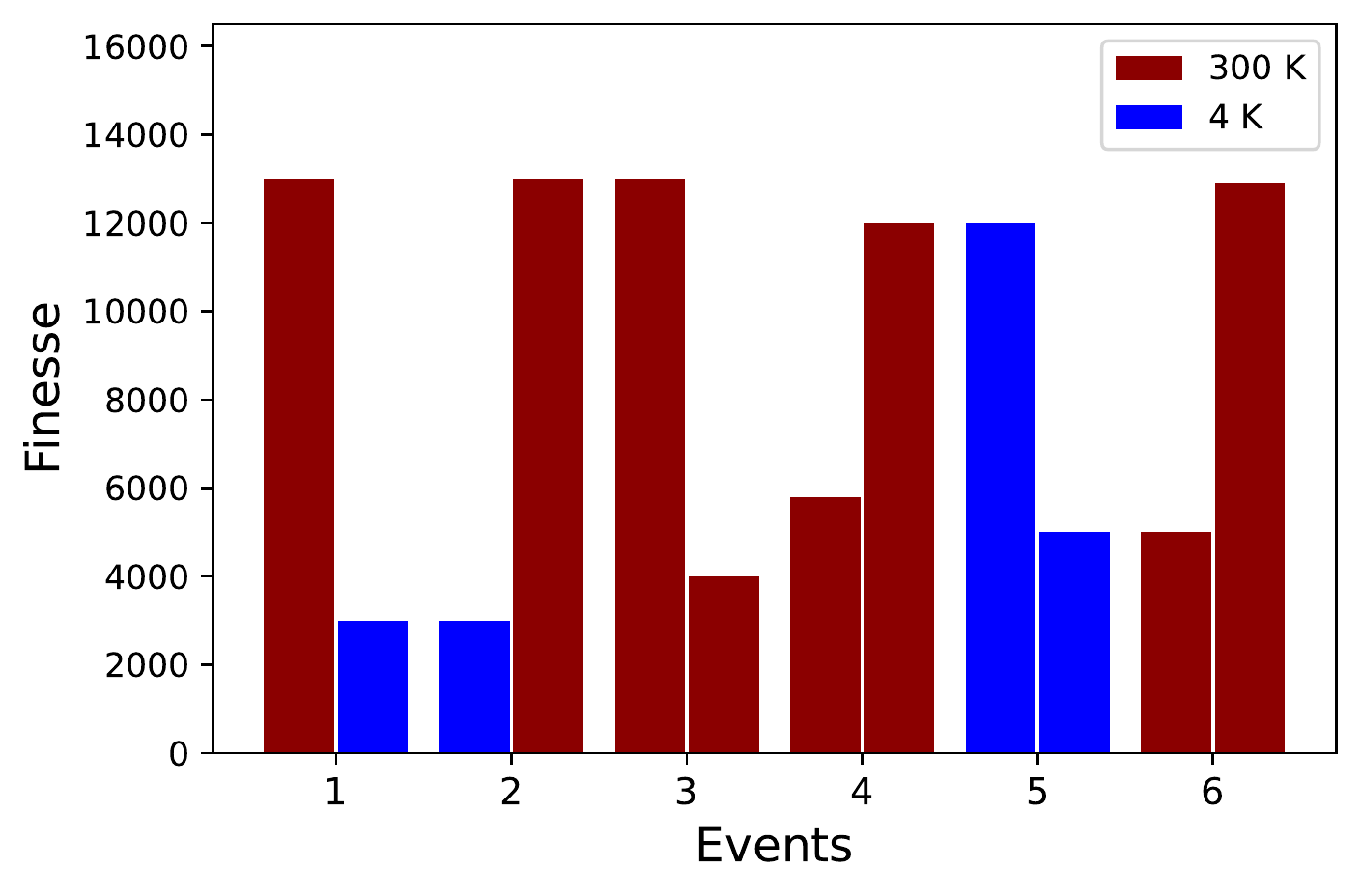}
\caption{\label{fig:finesse-events} Finesse changing caused by different events. The red and blue bars represent the finesse measurement at room temperature (300~K) and at cryogenic temperature (4~K), respectively.  
}
\end{figure}

Finesse preservation is essential to keep the power build-up high. In this section we report events that modified the cavity finesse. Finesse measurements followed the method described in section ~\ref{sec:cavity-characterization}. The measurements were performed at room temperature (300~K) and after the cryostat cooldown (4~K). The cavity finesse was preserved at its room temperature value after the cooldown of the cryostat on most cooldowns, however we observed some events that drastically modified the cavity finesse value. Fig. \ref{fig:finesse-events} shows the finesse changing occurred for six different events. In event 1, the finesse degraded from $F =$ 13000 to $F =$ 3000. On this occasion a valve which separates the cryogenic chamber from the turbo pump was left open during the cooldown of the cryostat. We guess that this led to a high vacuum pressure such that the condensation of gases on the cold cavity mirror might have resulted in contamination of the mirror coating. After warming up the cryostat (event 2) the finesse recovered its initial value of 13000. In event 3, the flat mirror was damaged when high circulating power was coupled into the cavity at room temperature (300~K). The estimated circulating power was around 70~MW/cm${^2}$ (1.7 W peak transmitted power) and the finesse dropped from $F =$ 13000 to $F =$ 4000. After that, the flat mirror was replaced by a new one. Initially, the cavity finesse with the new mirror was $F =$ 5800 at room temperature, but it increased to $F =$ 12000 after cleaning the flat mirror three times using a First Contact polymer and the finesse was preserved after the cooldown. With a transmitted power of 3.5~W (170~MW/cm$^{2}$) at cryogenic temperature (4~K), that was the record transmitted power observed in our system. However, this again led to damage of the flat mirror and the finesse dropped to $F =$ 5000 (event 5). The cavity finesse could be raised back up to 12900 by physically displacing the curved mirror and moving the cavity mode to a different spot on the flat mirror (event 6). This hints at mirror damage due to the high intensities. Table ~\ref{tab:finesse-changing} summaries the finesse changing events.

\begin{table}[htb!]
    \centering
        \begin{tabular}{| p{0.1\linewidth} | p{0.27\linewidth} | p{0.5\linewidth} |}
\hline  
Event & Finesse & Description \\
\hline
1 & 13k $\rightarrow$ 3$k$ & Cryostat cooldown with valve left open \\ \hline  
2 & 3k $\rightarrow $13$k$ & Finesse increased after warming up. \\ \hline 
3 & 13k $\rightarrow$ 4$k$ & Damage on the flat mirror after 1.7 W transmitted power. \\ \hline 
4 & 5.8k $\rightarrow$ 12$k$ & New flat mirror. Finesse incresead after 3 rounds of clean up. \\ \hline 
5 & 12k $\rightarrow$ 5$k$ & Damage on the flat mirror after 3.5 W transmitted power. \\ \hline 
6 & 5k $\rightarrow$ 12.9$k$ & Physically moved the curved mirror. \\ \hline 
    \end{tabular}
    \caption{Summary of events that modified the cavity finesse.}
    \label{tab:finesse-changing}
\end{table}

\subsection{Overlapping of cavity mode and ion position (Cryo cooldown)}

\begin{figure}[H]
\includegraphics[scale=0.66]{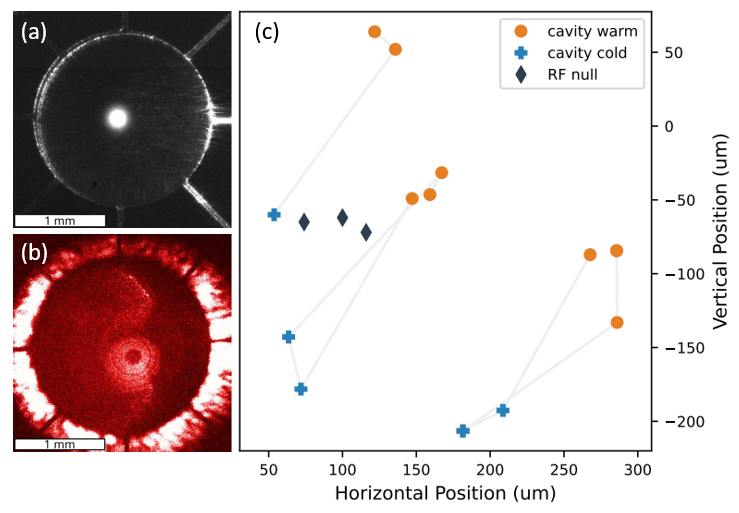}
\caption{\label{fig:cavity-displaciment} (a) cavity mode and front electrodes imaged through the cavity mirror. (b) ion and front electrode imaged through the asphere. The ion is out-of-focus as it was located 650 $\mu$m behind the electrode. The mirror in the asphere blocked light in a central cone, leading to the ring-like appearence of the ion. (c) Cavity displacement during the cooldown of the cryostat from the room temperature location (orange dots) to the 4 Kelvin location (blue crosses). Three independent reassemblies of the cavity mount is shown by the grouped gray lines.
}
\end{figure}

Overlapping the cavity mode to the ion position required knowing their relative positions in the radial plane of the ion trap. The cavity beam and the Mg$^+$ fluorescence do not share a common optical path because the flat cavity mirror coating blocks the ion fluorescence at 280 nm and the cavity beam is not refracted by the asphere due the central hole holding the mirror. In order to estimate their relative positions, we had to image the cavity mode and the ion separately, and to use the front electrode as a common reference to combine information from both images. Fig. \ref{fig:cavity-displaciment}(a) and (b) show the cavity mode imaged through the cavity mirror and the ion imaged through the asphere, respectively. Based on the images, we determine the cavity mode and ion position relative to the front electrode center. The front electrode center served as the origin position measurements.

Thermal distortions during the cooldown of the cryostat from room temperature to 4 K changes the size of parts of the cavity and trap mounts.

Fig. \ref{fig:cavity-displaciment}.(c) shows the cavity position mode location at room temperature (cavity warm) and at 4 K (cavity cold) for three independent mechanical reassemblies of the cavity mount. The cavity mode location was measured by imaging through the mirror in a $2f-2f$ configuration. We estimated that the precision of this method was on the order of $\pm$10 $\mu$m in both horizontal and vertical direction. Relative to the trap, the cavity mode repeatably moved roughly 100 $\mu$m down and 100 $\mu$m to the left after the cooldown cycles of the cryostat, and returned to the initial position after warming it up. Mechanical adjusts of the cavity and ion trap were only possible at room temperature. Once we had characterized the cooldown position shift, we were able to anticipate and compensate the cavity mode displacement. However, it remains unclear which part of the cavity or trap assembly is primarily responsible for the observed displacement of the cavity mode relative to the front electrode. The overall relative thermal expansion coefficient from room temperature to 4 K of titanium is $-171.9 \times 10^{-5}$.\cite{bradley2013properties} 
The side length of the cavity mount is 50 mm, which contracts by 85 $\mu$m. Moreover, the ion trap was mounted on a PCB made of FR-4, a multi-layer epoxy laminate material that might have deformed in complex ways. Finally, the asphere holding the flat cavity mirror might have contracted resulting in a tilt, which could have moved the cavity mode.

\begin{figure}[H]
\includegraphics[scale=0.52]{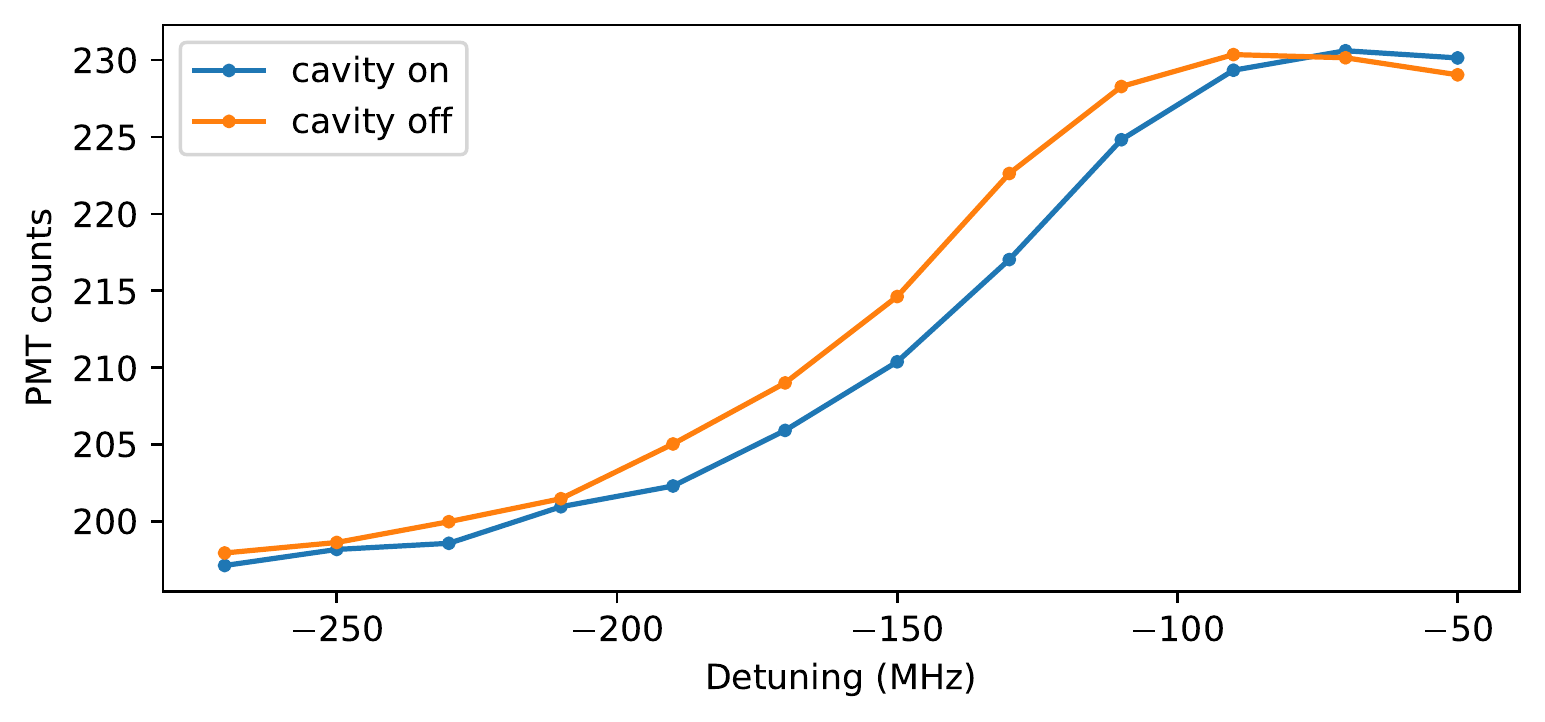}
\caption{\label{fig:lineshift-cavity} Observation of the atomic lineshift to overlap the ion position to the cavity mode location using a DC electric field. The ion was pushed out 250 $\mu$m, radial micromotion broadened the transition. A lineshift of 10~MHz was observed when the cavity was on with $\approx$25~MW/cm$^2$ circulating intensity. The calculated differential AC Stark shift at 25~MW/cm$^2$ is 4~MHz.
}
\end{figure}

Once the cryostat was cold, the position of the cavity mode was not tunable anymore. To improve the overlapping between cavity mode and the ion, we initially applied a DC electric field to adjust the ion position. The electric field pushed the ion out of the RF null toward the cavity mode. Consequently, the ion experienced micromotion which broadened the atomic transition due the Doppler effect. Figure \ref{fig:lineshift-cavity} shows the result of two laser frequency scans with the ion displaced 250 $\mu$m from the trap RF null toward the cavity mode. One was taken with no light in the cavity, and the other with a cavity peak circulating power of $\approx$25~MW/cm$^2$ (700 mW cavity transmitted power, 150 ppm per-mirror intensity transmission coefficient, 110 $\mu$m waist at the ion position). Only the red half of the resonance was measured because the frequency tuning was too slow to sample the blue side and recool quickly enough before the ion was heated out of the trap. At 1064~nm, the differencial AC-Stark shift on the 3$^{2}$S$_{1/2} \rightarrow 3^{2}$P$_{3/2}$ transition of Mg$^+$ is 0.16~Hz/(W/cm$^2$). The observed 10~MHz lineshift is on the same order of magnitude as the expected 4~MHz differential AC-Stark shift. It is also important to point out that the lineshift was only observed when the ion was positioned near to the calibrated cavity mode location. 
We therefore take this as evidence that the ion is in the cavity mode. However we have not been able to rule out that changes in the fluorescence profile due to shifts in micromotion of the ion may have produced this shift. 

\section{Conclusion}

In this work we have presented the design and performance of an optical build-up cavity integrated to a ring RF ion trap operating in a closed-cycle cryostat vacuum system with temperatures close to 4~K. We have overcome several technical challenges the setup presented. We have been able to suppress vibrations due the kicks of the closed-cycle cryostat and lock the 1064~nm laser into the cavity by suspending the cavity mount using copper braids. When we tried to run the cavity with high intracavity powers (more than 100~mW of input power) thermal cavity changes started to prevail in the system but we could still lock the cavity by setting VCO voltage  of the locking AOM to negative values avoiding the run-away process and by feeding back on the 1064~nm laser temperature to follow the cavity resonance. The maximal light intensity on the flat mirror we could achieve before it damages was 170~MW/cm$^{2}$ corresponding to an optical trap depth of 10~mK for Mg$^{+}$ ions in the trapping region. The ion could be overlapped to the cavity mode position by applying a DC electric field.

\begin{acknowledgments}

We acknowledge funding from Swiss National Science Foundation
under Grant No. BSCGI0 157834, and through the NCCR QSIT, a National Centre of
Competence in Research, Grant number 51NF40-185902.

\end{acknowledgments}

\nocite{*}
\bibliography{aipsamp}

\end{document}